\documentclass[12pt]{article}

\usepackage{amsmath, euscript, amssymb, amsfonts}
\usepackage[matrix,arrow,curve]{xy}



\newcommand{\oR}{{\mathbb R}}
\newcommand{\oV}{{\mathbb V}}
\newcommand{\oC}{{\mathbb C}}
\newcommand{\oT}{{\mathbb T}}

\newcommand{\oZ}{{\mathbb Z}}

\newcommand{\supp}{\mathop{\rm supp}\nolimits}

\renewcommand{\Re}{\mathop{\mathrm{Re}}\nolimits}
\renewcommand{\Im}{\mathop{\mathrm{Im}}\nolimits}

\begin{document}

\bigskip

\hfill FIAN-TD/2009-23

\bigskip

\baselineskip=20pt

\vspace{2cm}

\begin{center}

{\Large\bf Quantum field theory with a fundamental length:\\
A general mathematical framework}\footnote{\copyright 2009
American Institute of Physics. This  article appeared in J. Math.
Phys. 50, 123519 (2009) and may be found at
http://link.aip.org/link/?JMP/50/123519}

\vspace{1cm}

{\bf M.~A.~Soloviev}\footnote{E-mail: soloviev@lpi.ru}

\vspace{0.5cm}

 \centerline{\sl P.~N.~Lebedev Physical Institute}
 \centerline{\sl Russian Academy of Sciences}
 \centerline{\sl  Leninsky Prospect 53, Moscow 119991, Russia}

\vskip 3em

\end{center}

\begin{abstract}

We review and develop a  mathematical framework for  nonlocal
quantum field theory (QFT) with a fundamental length. As an
instructive example, we re\-examine the normal ordered Gaussian
function of a free field and find the primitive analyticity domain
of  its $n$-point vacuum expectation values. This domain  is
smaller than the usual future tube of local QFT, but we prove that
in difference variables, it has the same structure of a tube whose
base is the $(n-1)$-fold product of a Lorentz invariant region. It
follows that this model satisfies Wightman-type axioms with an
exponential high-energy bound which does not depend on $n$,
contrary to the claims  in the literature. In our setting, the
Wightman generalized functions are defined on test functions
analytic in the complex $l$-neighborhood of the real space, where
$l$ is an $n$-independent constant playing the role of a
fundamental length, and the causality condition is formulated with
the use of an analogous function space associated with the light
cone. In contrast to the scheme proposed by Br\"uning and
Nagamachi [J. Math. Phys. 45 (2004) 2199] in terms of
ultra-hyperfunctions, the presented theory obviously becomes local
as $l$ tends to zero.

\end{abstract}

\newpage
\vskip 2em

\section{\large Introduction}

Since the seminal works of Jaffe~\cite{Jaffe} and Meyman~\cite{M},
it has been well recognized that the localizability condition
imposes restrictions on the high-energy behavior of quantum
fields. Specifically,  the momentum-space expectation values of
local fields have less than exponential growth. This has suggested
a possible way of constructing a self-consistent theory of
nonlocal interactions by going beyond this boundary. At the
present time, nonlocal field theories of this kind continue to
attract interest because of their interplay with string theory and
holographic models, see, e.g.,~\cite{FS2,G,K,HRR} for a discussion
of this and related issues. It is also important that these
theories provide an alternative to the use of the idealized
concept of local commutativity ( also called  microcausality)  and
are aimed at finding a more physically motivated implementation of
causality.

Iofa and Fainberg~\cite{IF1,IF2} were first to show that the
Wightman axiomatic approach~\cite{SW} can be extended to nonlocal
field theories with an exponential growth of the off-shell-mass
amplitudes. In this case, a field must be averaged with analytic
test functions in order to yield a well defined operator, and this
makes the formulation~\cite{SW} of local commutativity impossible
because it uses test functions of compact support.
In~\cite{IF1,IF2}, the microcausality axiom was replaced (for the
case of a single scalar field) by the condition of symmetry of the
analytic Wightman functions under the permutations of their
arguments. Subsequently it was shown~\cite{FS1} that a natural way
of formulating causality in nonlocal field theories with analytic
test functions is by using a suitably adapted notion of carrier of
an analytic functional. In~\cite{FS1}, it was also proved that
such a formulation implies the symmetry of the  Wightman functions
when their domain of analyticity is nonempty.

In~\cite{IF1,IF2,FS1}, it was assumed that  the $n$-point vacuum
expectation values of nonlocal fields grow in momentum space not
faster than
\begin{equation}
\exp\left\{\ell\sum_{j=1}^{n-1}|q_j|\right\}, \label{1.1}
 \end{equation}
where $q_j$ is the momentum conjugate to the relative coordinate
$\xi_j=x_j-x_{j+1}$ and $\ell$ is an $n$-independent constant
playing the role of a fundamental length. This assumption is
equivalent to saying that, in the coordinate representation, the
vacuum expectation values, considered as generalized functions of
the variables  $\xi_j$, are  defined on the test function space
$A_\ell(\oR^{4(n-1)})$ whose elements are analytic in the complex
$\ell$-neighborhood of the real space $\oR^{4(n-1)}$ and decrease
rapidly at infinity. (The exact definition of this space is given
in Sec.~III.) More recently, Br\"uning and Nagamachi~\cite{BN}
proposed to replace the uniform bound~\eqref{1.1} with a weaker
condition. Simply stated, this condition means that the growth of
the vacuum expectation values in every variable $q_j$ at fixed
$q_i$, $i\ne j$, is not faster than $\exp(\ell|q_j|)$. The formal
definition~\cite{BN} is given  in terms of a special class of
generalized functions called ultra-hyperfunctions.  It can be
rewritten in our notation as the requirement that the $n$-point
vacuum expectation value considered as a generalized function of
the  $\xi_j$'s is defined on each of the spaces
\begin{equation}
A_\infty(\oR^{4(j-1)})\otimes A_\ell(\oR^4)\otimes
A_\infty(\oR^{4(n-j-1)}),\quad j=1,\dots, n-1 .
 \label{1.2}
 \end{equation}
Br\"uning and Nagamachi claimed that such a modification is
necessary to make the theory applicable to the nonlocal model
$:e^{ g\phi^2}\!\!:(x)$, where $\phi$ is a free scalar field and
$:\,:$ stands for the normal ordering.

In the present paper, we show that this claim results from an
imperfect description of the analyticity domain of the
corresponding Wightman functions. We prove that the toy model
$:e^{ g\phi^2}\!\!:(x)$ completely fits in the original
framework~\cite{IF1,IF2,FS1}. So, this example does not give any
grounds for a more complicated scheme with mixed-type spaces. Our
main aim here is to present a development of the ideas proposed
in~\cite{IF1,IF2,FS1} and give the rigorous and up-to-date
formulation of a general mathematical framework for treating
quantum field models with a fundamental length. We also show that
the presented theory becomes local in the limit $\ell\to0$,
whereas the existence of a local limit for the modified
scheme~\cite{BN} is problematic.

In Sec.~II, we describe exactly the analyticity domains of the
Wightman functions of the field  $:e^{ g\phi^2}\!\!:$ and show
that these domains are considerably larger than those found
in~\cite{BN} and have a simpler structure. Making use of this
result, we prove in Sec.~III that the  $n$-point vacuum
expectation values of this field, when considered as a generalized
function of the difference variables $\xi_j$, are well defined on
the spaces $A_\ell(\oR^{4(n-1)})$, where $\ell=\sqrt{g/6}$. In
Sec.~IV, we show that these generalized functions  satisfy the
quasilocality condition introduced in~\cite{FS1}. A general
mathematical framework for quantum field theory (QFT) with a
fundamental length is discussed in detail in Sec.~V, where we also
compare our approach with that of Ref.~\cite{BN} and reveal some
essential differences in their consequences. Sec.~VI contains
concluding remarks.

\section{\large Wightman functions of the field $:\exp g\phi^2:(x)$}

Let  $\phi$ be a free neutral scalar field of mass $\mu\geq 0$ in
Minkowski space. We recall some simple but important properties of
its two-point vacuum expectation value
\begin{equation}
\langle\Psi_0,\phi(x_1)\phi(x_2)\Psi_0\rangle= \Delta_+(x_1-x_2)=
\frac{1}{(2\pi)^3} \int \theta(p^0) \delta (p^2-\mu^2)
e^{-ip\cdot(x_1-x_2)}{\rm d}p.
 \label{2.1}
 \end{equation}
The function  $\Delta_+(z)$, $z=x+iy$, is analytic in the domain
$\oR^4+i\oV^-$, where $\oV^-$ is the lower light cone   $\{y\in
\oR^4\colon y^2=(y^0)^2-\mathbf y^2>0,\, y^0<0\}$, and satisfies
the inequality
  \begin{equation}
|\Delta_+(x+iy)|\leq \frac{1}{4\pi^2}\cdot\frac{1}{y^2},\qquad
y\in\oV^-.
   \label{2.2}
 \end{equation}
Indeed,  this function is Lorentz invariant and hence  we can
assume, without loss of generality, that $y=(-\tau, 0,0,0)$,
$\tau>0$. Then,
\begin{multline}
|\Delta_+(x^0-i\tau,{\bf x})|=\frac{1}{(2\pi)^3}
\left|\int\frac{e^{-i\omega({\bf p})(x^0-i\tau)}} {2\omega({\bf
p})}e^{i{\bf px}}{\rm d}{\bf p}\right|\\\leq
 \frac{1}{4\pi^{2}}\int_0^\infty \frac{e^{-\tau \sqrt{s^2+\mu^2}}}{\sqrt{s^2+\mu^2}}
\,s^{2}{\rm d} s \leq
 \frac{1}{4\pi^2} \int_0^\infty e^{-\tau s}
s{\rm d} s=\frac{1}{4\pi^2}\cdot\frac{1}{\tau^2}. \notag
\end{multline}
We will use the estimate~\eqref{2.2} to describe properties of the
vacuum expectation values of the field
\begin{equation}
:\exp
g\phi^2:(x)=\sum_{r=0}^\infty\frac{g^r}{r!}:\!\phi^{2r}\!:(x).
 \label{2.3}
 \end{equation}
As shown by Rieckers~\cite{R}, the field~\eqref{2.3} can be
implemented in the Hilbert space of $\phi$ as an operator-valued
generalized function with test functions in any Gelfand-Shilov
space $S^\alpha$, where $\alpha<1$. Rieckers  argued that this
field obeys all Wightman axioms, except for locality, and noted
that its two-point function  is well defined even on a larger
space $S^{1,\lambda}$, where $\lambda$ depends on $g$. We shall
consider the $n$-point vacuum expectation values
\begin{equation}
{\mathcal W}_n(x_1,\dots x_n)=\langle\Psi_0, :\exp
g\phi^2:(x_1)\,\ldots :\exp g\phi^2:(x_n) \Psi_0\rangle,
\label{2.4}
\end{equation}
where $n\geq2$. We first find the analyticity domains of the
corresponding Wightman functions $W(\xi_1,\dots\xi_{n-1})$
depending on the relative coordinates $\xi_j=x_j-x_{j+1}$. These
functions exist by the translation invariance of~\eqref{2.4} and
satisfy
\begin{equation}
W(x_1-x_2,\dots,x_n-x_{n-1})={\mathcal W}_n(x_1,\dots x_n).
 \label{2.*}
\end{equation}
We let $\zeta_j=\xi_j+i\eta_j$ denote the complex difference
variables and use the notation
\begin{equation}
\oV^-_l=\{\eta\in \oR^4\colon \eta^2>l^2, \, \eta^0<0\}.
 \label{2.5}
\end{equation}

\medskip

{\bf Theorem~1.} {\it
 The Wightman function $W(\zeta_1,\dots \zeta_{n-1})$ of the field $:\exp g\phi^2:$
 is analytic in the tube $(\oR^4+i\oV^-_\ell)^{n-1}$,
where $\ell=\sqrt{g/6}$.

\medskip

Proof.} From the Wick theorem it follows that the $n$-point vacuum
expectation values of any normal ordered entire function of the
free field $\phi$ are representable as formal power series in the
variables
\begin{equation}
w_{ij}=\Delta_+(x_i-x_j), \qquad 1\leq i<j\leq n.
 \label{*}
\end{equation}
Let  $R$ be a multi-index whose components $r_{ij}$,  $1\leq
i<j\leq n$, ranges over the set  $\oZ_+$ of nonnegative integers.
If
\begin{equation}
\varphi(x)=\sum_{r=0}^\infty\frac{d_r}{r!}:\phi^r:(x),
 \label{2**}
\end{equation}
then the formal power expansion  has the form
\begin{equation}
\langle \Psi_0,\, \varphi(x_1)\ldots \varphi(x_n)\Psi_0\rangle
=\sum_R \frac{D_R}{R!}\,w^R,
 \label{2.6}
\end{equation}
where
$$
w^R=\prod_{1\leq i<j\leq n}w_{ij}^{r_{ij}},\qquad R!=\prod_{1\leq
i<j\leq n}r_{ij}!.
$$
An analysis carried out by Jaffe~\cite{Jaffe2} shows that the
coefficients $D_R$ in~\eqref{2.6} are related to the initial
coefficients $d_r$  by
\begin{equation}
D_R=\prod_{j=1}^nd_{R_j},
\label{2.7}
\end{equation}
 where
$R_j=r_{1j}+\ldots+r_{j-1,j}+r_{j,j+1}+\ldots+r_{jn}$ is the
occurrence number of $x_j$ in the monomial $w^R$.

We consider the terms of the series on the right-hand side
of~\eqref{2.6} as functions of the complex variables
$\zeta_j=\xi_j+i\eta_j$ lying in the domain $\oR^4+i\oV^-$, where
every monomial $w^R$ is analytic. In the case under study
$$
d_r=\begin{cases} g^{r/2}\dfrac{r!}{(r/2)!}&\text{for even $r$,}\\
0 &\text{for odd $r$,}
\end{cases}
$$
and by the Stirling formula these coefficients satisfy
\begin{equation}
d_r^2\leq  (2g)^r r!.
 \label{2.8}
\end{equation}
 Now we define   $w_{ij}$ and $r_{ij}$ for $i>j$ by setting
 $w_{ij}=w_{ji}$ and  $r_{ij}=r_{ji}$. We also set
 $r_{jj}=0$ for all $j=1,\dots,n$.  Note that then
$R_j=\sum\limits_{i=1}^n r_{ij}$  is the sum of elements of the
$j$-th column of the symmetric matrix  $(r_{ij})$. Moreover,
 $$
  \left|\frac{w^R}{R!}\right|^2=\prod_{\substack{i,j=1\\i\ne j}}^n
 \frac{|w_{ij}|^{r_{ij}}}{r_{ij}!}.
 $$
 Using~\eqref{2.7}, \eqref{2.8} and taking into account that
$R_j!/(r_{1j}!\dots r_{nj}!)$ is a multinomial coefficient, we
obtain
\begin{equation}
\left|\frac{D_R}{R!}\,w^R\right|^2\leq
\prod_{j=1}^n\left((2g)^{R_j}R_j!\prod_{\substack{i=1\\i\ne j}}^n
\frac{|w_{ij}|^{r_{ij}}}{r_{ij}!} \right)\leq
(2g)^{2|R|}\prod_{j=1}^n\left(\sum_{\substack{i=1\\i\ne j}}^n
|w_{ij}|\right)^{R_j},
 \label{2.9}
\end{equation}
where $2|R|=\sum_jR_j$. We note that if all the vectors
$\Im\zeta_j$ belong to $\oV^-_l$ , where $l>0$, then the sum on
the right-hand side of~\eqref{2.9} is uniformly bounded by a
constant\footnote{This point seems to be overlooked in~\cite{BN}.}
independent of $n$. Indeed, $z_i-z_j=\zeta_i+\dots+\zeta_{j-1}$
for any $j>i+1$ and hence
$$
\Im(z_i-z_j)\in (j-i)\oV^-_l=\oV^-_{(j-i)l} \qquad \text{for all
$j> i$},
 $$
because the set $\oV^-_l$ is convex. From~\eqref{2.2} it follows
that for such arguments,
\begin{equation}
|\Delta_+(z_i-z_j)|\leq \frac{1}{4\pi^2
l^2}\cdot\frac{1}{(j-i)^2}, \qquad j>i.
 \label{2.9*}
\end{equation}
Therefore,
\begin{equation} \sum_{\substack{i=1\\i\ne j}}^n
|w_{ij}| \equiv \sum_{i=1}^{j-1} |\Delta_+(z_i-z_j)|
+\sum_{i=j+1}^n |\Delta_+(z_j-z_i)|\leq \frac{1}{2\pi^2
l^2}\sum_{k=1}^{n-1}\frac{1}{k^2}
 \label{2.10}
\end{equation}
everywhere in the specified region of the complex variables.
Combining~\eqref{2.9} and \eqref{2.10} and taking into account
that $\sum_{k=1}^\infty(1/ k^2)=\pi^2/6$, we obtain
\begin{equation}
\left|\frac{D_R}{R!}\,w^R\right|\leq \left(\frac{g}{\pi^2
l^2}\sum_{k=1}^{n-1}\frac{1}{k^2}\right)^{|R|}<\left(\frac{g}{6
l^2}\right)^{|R|}.
 \label{2.11}
\end{equation}
We conclude that the formal  series representing the Wightman
functions $W(\zeta_1, \dots,\zeta_{n-1})$ converge absolutely if
the imaginary parts of all $\zeta_j$ belong to $\oV^-_l$, where
$l>\ell=\sqrt{g/6}$. The convergence is uniform in this region and
hence the limit functions are analytic in
$(\oR^4+i\oV^-_\ell)^{n-1}$. The theorem is proved.

\medskip

{\large Remark~1.}  In the simples case of zero mass, it is easy
to obtain an explicit expression for the two-point function of the
field~\eqref{2.3}. Indeed, we have
    \begin{equation}
    \langle \Psi_0,\, :\exp
  g\phi^2:(x_1) :\exp
  g\phi^2:(x_2)\Psi_0\rangle
     =  \sum_{m=0}^\infty  \frac{(2m)!}{(m!)^2}(g\Delta_+(x_1-x_2))^{2m}.
  \label{2.12}
  \end{equation}
  If $z\in \oC$ and $4|z|<1$, then
   $$ \sum_{m=0}^\infty
  \frac{(2m)!}{(m!)^2}z^{m}=\frac{1}{\sqrt{1-4z}}\, .
   $$
Furthermore, if  $\mu=0$, then
$\Delta_+(\zeta)=-\dfrac{1}{4\pi^2}\cdot\dfrac{1}{\zeta^2}$ and
the estimate~\eqref{2.2} is exact. Considering that
$|\zeta^2|\geq(\Im \zeta)^2$ for all $\Im \zeta\in \oV^-$, we
deduce that in this case, the two-point function~\eqref{2.12} has
the form
$$
\frac{1}{\sqrt{1-4g^2\Delta_+(\zeta)^2}}=
  \frac{\zeta^2}{\sqrt{(\zeta^2)^2-\ell_0^4}},\qquad \mbox{where} \quad
  \ell_0=\dfrac{1}{\pi}\sqrt{\dfrac{g}{2}},
$$
and its primitive analyticity domain is the tube
$\oR^4+i\oV^-_{\ell_0}$. Note that
$\ell=(\pi/\sqrt{3})\ell_0\,\approx 1.8\,\ell_0$.

\medskip

{\large Remark~2.} The analyticity domain found for
$W(\zeta_1,\dots \zeta_{n-1})$ in~\cite{BN}, p.~2225, has the form
$$
\bigcup_{i=1}^{n-1} {\mathcal V}_{R,\epsilon,i},
$$
where
$$
{\mathcal V}_{R,\epsilon,i}=\{\zeta\in\oC^{4(n-1)}\colon
-\Im\zeta_i\in\oV_+ +(\ell_0+\epsilon,{\bf0}), -\Im\zeta_j\in
\oV_+ +(R,{\bf0}), j\ne i\}
$$
and $\epsilon>0$ can be taken arbitrarily small, but  $R$ is an
undetermined  constant depending on~$\epsilon$.

\section{\large The appropriate test functions}

As a simple application of the above theorem, we  specify the test
function spaces that are adequate to the generalized
functions~\eqref{2.4}. Let $\ell$ be a positive number or $\infty$
and let $|y|=\max_ {1\leq j\leq d}|y_j|$. We denote by
$A_\ell(\oR^d)$ the topological vector space of  functions
analytic in the tubular domain $\{x+iy\in \oC^d\colon |y|<\ell\}$
and such that all the norms
 \begin{equation} \|f\|_{l,N}=\sup_{|y|\leq
 l}\sup_{x\in\oR^d} (1+|x|)^N|f(x+iy)|,\qquad l<\ell,\,\,
 N=0,1,2,\ldots,
     \label{3.1}
     \end{equation}
     are finite. Let
$\lambda=1/(e\ell)$ and $\lambda=0$ for the special case
$\ell=\infty$. The space $A_\ell(\oR^d)$ coincides with the space
$S^{1, \lambda}(\oR^d)$ which by definition~\cite{GS2} consists of
all infinitely differentiable functions on $\oR^d$ with the
property that, for each $\bar{\lambda}>\lambda$,
\begin{equation}  (1+|x|)^N|\partial^\kappa
f(x)|<C_{N\bar{\lambda}}\bar{\lambda}^{|\kappa|}\kappa^\kappa,
 \label{3.1*}
     \end{equation}
where $N$ ranges over $\oZ_+$,  the multi-index $\kappa$ ranges
over $\oZ_+^d$, and  $C_{N\bar{\lambda}}$ is a constant depending
on $f$. As shown in~\cite{GS2}, the Taylor series expansion of any
function belonging to $S^{1, \lambda}(\oR^d)$ defines an element
of $A_\ell(\oR^d)$. Conversely, using  Cauchy's integral formula,
one can easily verify that the restriction of any element of
$A_\ell(\oR^d)$ to the real space $\oR^d$ belongs to $S^{1,
\lambda}(\oR^d)$ and the map $A_\ell(\oR^d)\to S^{1,
\lambda}(\oR^d)$ is continuous. In what follows, the notation
$A_\ell(\oR^d)$ is more convenient for brevity.\footnote{In
Ref.~\cite{FS1}, this space was denoted by $\mathfrak
A_\ell(\oR^d)$.}

\medskip

{\large Remark~3.} The space
$A_\infty(\oR^d)=\bigcap_{\lambda>0}S^{1, \lambda}(\oR^d)$
consists of the entire analytic functions that decrease faster
than any inverse  power of $|x|$ as $|x|\to\infty$ and it
coincides with the space $\mathcal T(T(\oR^d))$ in the notation
used in~\cite{BN}. The elements of its dual space $\mathcal
T(T(\oR^d))'=A'_\infty(\oR^d)$ are said to be tempered
ultra-hyperfunctions. The space $A_\ell(\oR^d)$ was denoted
in~\cite{BN} by $\mathcal T(T(O)$, where $T(O)=\oR^d+iO$ and
$O=(-\ell, \ell)^d$.

\medskip

According to Gelfand and Shilov~\cite{GS2},  $S^{1,
\lambda}(\oR^d)=A_\ell(\oR^d)$ is a complete metrizable nuclear
space. Therefore, it belongs to the class FS of Fr\'echet-Schwartz
spaces. In particular, $A_\ell(\oR^d)$ is a Montel space, and
hence, reflexive.  These facts can be used to derive its other
properties useful for applications. As a simple example, we prove
in  Appendix that the space $S^0(\oR^d)$, which is the Fourier
transform of the space $\mathcal D(\oR^n)$ of smooth functions of
compact support, is dense in  $A_\ell(\oR^d)$ for any $\ell$. A
fortiori, $A_\infty(\oR^d)$ is dense in $A_\ell(\oR^d)$. In the
textbook~\cite{GS2} it is shown that the Fourier transformation
$f(x)\to\hat f(p)=\int f(x)e^{ipx}{\rm}dx$ is an isomorphism of
$S^{1, \lambda}(\oR^d)$ onto the space $S_{1, \lambda}(\oR^d)$
consisting of all smooth functions $g(p)$ with the finite norms
\begin{equation}
\|g\|'_{l,N}=\max_{|\kappa|\leq N}\sup_{p\in
\oR^d}|\partial^\kappa g(p)|
\exp\left\{l\sum\limits_{j=1}^d|p_j|\right\}, \qquad
l<1/(e\lambda).
 \label{3.2}
 \end{equation}

 \medskip

{\bf Theorem~2.} {\it Considered as a generalized function of the
difference variables, the $n$-point vacuum expectation value of
$:\exp g\phi^2:$ is well defined on the test function space
$A_\ell(\oR^{4(n-1)})$, where $\ell=\sqrt{g/6}$.

\medskip

Proof.} The tempered distribution $w^R(\xi)$ in~\eqref{2.6} is the
boundary value of the function $w^R(\xi+i\eta)$ analytic in the
tube $(\oR^4+i\oV^-)^{n-1}$. Therefore, for each $\eta\in
(\oV^-)^{n-1}$ such that $|\eta|<\ell$, we have
 \begin{equation}
     (w^R,f)=\int w^R(\xi+i\eta)\,f(\xi+i\eta)\, {\rm
  d}\xi,\quad f\in A_\ell(\oR^{4(n-1)}).
  \label{3.3}
 \end{equation}
Let $\eta_j=(-l, 0,0,0)$ for all $j$. From~\eqref{2.11} and the
definition~\eqref{3.1}, it follows that
\begin{equation}
\left|\frac{D_R}{R!}\,(w^R,f)\right| \leq
C\,\|f\|_{l,4n-3}\left[\frac{g}{\pi^2
l^2}\sum_{k=1}^{n-1}\frac{1}{k^2}\right]^{|R|}.
  \label{3.5}
  \end{equation}
where $C=\int_{\oR^{4(n-1)}}(1+|\xi|)^{-4n+3}{\rm d}\xi$. Because
 $\sum_{k=1}^{n-1}(1/ k^2)$ is strictly less than $\pi^2/6$, there
exists an $l<\ell$ such that the number in the square brackets is
less than 1. Hence the number series $\sum_R (D_R/R!)(w^R,f)$
converges absolutely for each $f\in A_\ell(\oR^{n-1})$ and defines
a linear functional on  $A_\ell(\oR^{n-1})$ whose continuity is
ensured  by the factor $\|f\|_{l,4n-3}$ on the right-hand side
of~\eqref{3.5}. The theorem is proved.

\medskip

{\large Corollary.} {\it The generalized function~\eqref{2.4} is
well defined  on the space $A_{\ell_n}(\oR^{4n})$, where
$\ell_n=\ell(n-1)/2$.

\medskip

Proof.} We let $t$ denote the linear  transformation
\begin{equation}
(x_1,\dots, x_n)\longrightarrow (X,\xi_1,\dots \xi_{n-1}),
\label{3.6}
\end{equation}
where
$$
X=\frac{1}{n}(x_1+\dots+ x_n),\qquad \xi_j=x_j-x_{j+1}.
$$
Then, we have
$$
({\mathcal W}_n, f)=(W, f_t),\qquad f_t(\xi)=\int_{\oR^4}
f(t^{-1}(\xi,X))\,{\rm d}X,
 \notag
 $$
which is  a rigorous form of the formal relation~\eqref{2.*}. The
inverse transformation $t^{-1}$ is written as
\begin{equation}
 x_j=X-\frac{1}{n}\sum_{m=1}^{j-1}m\xi_m+
\frac{1}{n}\sum_{m=1}^{n-j}m\xi_{n-m}.
 \label{3.7}
\end{equation}
If the $\xi_j$'s in~\eqref{3.7} are changed for $\xi_j+i\eta_j$,
where $|\eta_j|<\ell$, then the variables $x_j$ gain imaginary
parts $y_j$ satisfying
\begin{equation}
|y_j|<\frac{\ell}{n}\left(\sum_{m=1}^{j-1}m+
\sum_{m=1}^{n-j}m\right)\leq
\frac{\ell}{n}\sum_{m=1}^{n-1}m=\ell\frac{n-1}{2}.
 \label{3.8}
\end{equation}
Hence, if $f\in A_{\ell_n}(\oR^{4n})$, then $f_t$ is analytic in
the domain $\{\xi+i\eta\in \oC^{4(n-1)}\colon |\eta|<\ell\}$. Let
 $l<\ell$. Using the inequalities  $|\xi|\leq
2|x|$ and $|X|\leq|x|$, we get
\begin{multline}
\|f_t\|_{l,N}\leq\sup_{|\eta|<l}\sup_\xi
(1+|\xi|)^N\int_{\oR^4}|f(t^{-1}(\xi+i\eta,X))|{\rm d}X\\
\leq C\sup_{|y|<l(n-1)/2}\sup_x(1+|x|)^{N+5}|f(x+iy)|
=C\,\|f\|_{l(n-1)/2,N+5},
 \label{3.9}
\end{multline}
where $C=2^N\int_{\oR^4}(1+|X|)^{-5}{\rm d}X$. Therefore, $f_t\in
A_\ell(\oR^{4(n-1)})$  and the map $A_{\ell_n}(\oR^{4n})\to
A_\ell(\oR^{4(n-1)})\colon f\to f_t$ is continuous.

\medskip

 {\large Remark~4.} The formulas~\eqref{3.7} and \eqref{3.8}
exhibit a simple geometric fact. Namely, if we have an ordered set
of points in $\oR^d$ and the spacing between adjacent points is
not greater than $\ell$, then the distance of any point from the
geometric center of the set does not exceed $\ell(n-1)/2$.

\section{\large Quasilocality}

The microcausality axiom~\cite{SW} expresses the idea of
independence of field measurements performed at spacetime points
separated by a spacelike interval. Explicitly,
$$
[\varphi(f), \psi(g)]\Psi=0
$$
for any observable fields $\varphi$ and $\psi$, for all states
$\Psi$ in their common domain and for any test functions $f$, $g$
whose supports are spacelike to each other. In the simplest case
of a scalar local field $\varphi(x)$ this amounts to saying that
for any $k\leq n$, the support of
 \begin{multline}
\langle \Psi_0,\,
\varphi(x_1)\dots\varphi(x_k)\varphi(x_{k+1})\dots\varphi(x_n)\Psi_0\rangle\\
 - \langle \Psi_0,\,
\varphi(x_1)\dots\varphi(x_{k+1})\varphi(x_{k})\dots\varphi(x_n)\Psi_0\rangle
  \label{4.1}
  \end{multline}
is contained in the wedge
\begin{equation}
\bar V_{(k,k+1)}=\{x\in\oR^{4n}\colon (x_k-x_{k+1})^2\geq 0\}.
  \label{4.2}
  \end{equation}
In~\cite{FS1}, it was argued that a natural generalization of this
axiom to the nonlocal QFT with test functions in $A_\ell(\oR^4)$
is the condition of continuity of the functional~\eqref{4.1} in
the topology of an analogous space associated with the
wedge~\eqref{4.2}. We recall the definitions introduced there.

For each set $O\subset\oR^d$, we define a space  $A_\ell(O)$ in
the following way. Let $\tilde O^l$ be the complex
$l$-neighborhood of $O$, consisting of those points $z\in \oC^d$
for which there is  $x\in O$ such that
$|z-x|\equiv\max_j|z_j-x_j|<l$. The space $A_\ell(O)$ consists of
all functions analytic in ${\tilde O}^\ell$ for which the norms
\begin{equation}
\|f\|_{O, l, N}=\max_{|\kappa|\leq N}\sup_{z\in \tilde O^l}
 |z^\kappa f(z)|,\qquad l<\ell,  \quad N=0,1,2,\dots,
\label{4.3}
     \end{equation}
are finite. We note that  $A_\ell(O)$  coincides with $A_\ell(\bar
O)$, where   $\bar O$ is the  closure of $O$. Using the relations
\begin{equation}
\max_{\kappa\leq N}|z^\kappa|=\max(1,|z|^N),\qquad (1+|z|)^N\leq
2^N\max(1, |z|^N),
 \notag
\end{equation}
it is easy to verify that in the particular case $O=\oR^d$, this
definition is equivalent to the definition of $A_\ell(\oR^d)$
given above. If $O_2\subset O_1$, then there is a natural
continuous injective map $A_\ell(O_1)\to A_\ell(O_2)$ called
``restriction morphism''. Let $f|O_2$ be the image of $f$ under
this morphism. Then  the relation $f|O_3=(f|O_2)|O_3$ holds for
any  $O_3\subset O_2\subset O_1$ and so we have a presheaf of
vector spaces.

Now we shall show that the vacuum expectation values of the field
~\eqref{2.3}, considered as generalized functions of the
difference variables $\xi_j=x_j-x_{j+1}$, satisfy the
quasilocality condition introduced in~\cite{FS1}. The proof
proceeds in the same manner as the derivation  of an analogous
theorem~\cite{SS}  for those Wick power series whose limits are
 defined on the space $S^1=\bigcup_{\lambda>0}S^{1,\lambda}$
and are localizable in the sense of
hyperfunctions.\footnote{In~\cite{SS}, a more complicated case of
essentially nonlocalizable infinite series of Wick powers is also
considered and their limits are shown to satisfy the asymptotic
commutativity condition introduced in~\cite{FS2}.}  In this
derivation, a key role is played by the Bargmann-Hall-Wightman
theorem~\cite{SW} which shows that the function $\Delta_+(z)$ can
be analytically continued to an extended tube $\oT^{\rm ext}$.
This tube is obtained by applying all complex Lorentz
transformations to the primitive analyticity domain
$\oR^4+i\oV^-$. The continued function is invariant under the
complex Lorentz group $L_+(\oC)$ and, in particular, under the
total reflection $z\to-z$. We note that the transposition $x_k$
and $x_{k+1}$ induces the transformation
\begin{equation}
\xi_{k-1}\to \xi_{k-1}+\xi_k,\quad \xi_k\to -\xi_k\quad
\xi_{k+1}\to \xi_{k+1}+\xi_k
 \label{4.3*}
\end{equation}
of the difference variables. (Strictly speaking, here we mean
that $n\ne 2$ and $1< k< n-1$. For $n= 2$ we have simply the
reflection $\xi_1\to-\xi_1$. For  $n>2$ and $k=1$, $k=n-1$ we
have, respectively, $\xi_1\to-\xi_1$, $\xi_2\to\xi_2+\xi_3$ and
$\xi_{n-2}\to \xi_{n-2}+\xi_{n-1}$, $\xi_{n-1}\to-\xi_{n-1}$.) The
functional~\eqref{4.1} written in these variables looks as follows
\begin{multline}
W(\xi_1,\dots,\xi_{k-1},\xi_k,\xi_{k+1},\dots, \xi_{n-1}) -\\
W(\xi_1,\dots,\xi_{k-1}+\xi_k,-\xi_k,\xi_k+\xi_{k+1},\dots,
\xi_{n-1}).
 \label{4.4}
     \end{multline}
Below we  use an extension of the  estimate~\eqref{2.2} to
$\oT^{\rm ext}$.

\medskip

{\bf Lemma.} {\it  For any $l>0$, the function $\Delta_+(z)$ is
analytic in the domain $\{z=x+iy\in\oC^4\colon x^2<-l^2< y^2\}$
and satisfies the inequality
\begin{equation}
|\Delta_+(x+iy)|\leq \frac{1}{2\pi^2}\cdot\frac{1}{y^2-x^2}
  \label{4.5}
  \end{equation}
everywhere in this domain.

\medskip

  Proof.} Suppose first that $y^2>0$. Because
 $\Delta_+$ is Lorentz invariant, we may assume
without loss of generality that the vectors $y$ and $x$ have the
form $y=(y^0,0,0,0)$ and $x=(x^0,x^1,0,0)$, where $x^1<0$. This
can be achieved by a boost with a subsequent rotation. Next we use
the invariance under $L_+(\oC)$ and apply the transformation
$\Lambda\colon (z^0, z^1,z^2, z^3)\to (iz^1, iz^0,z^2,z^3)$
belonging to this group. The vector $\Im\Lambda z=(x^1,x^0,0,0)$
lies in $\oV^-$ and $(\Im\Lambda z)^2=-x^2$. Using~\eqref{2.2}, we
obtain the inequality $4\pi^2|\Delta_+(x+iy)|\leq
\min[1/y^2,1/(-x^2)]$, which implies~\eqref{4.5}. Now let $y^2<
0$. Then, we may assume that $y=(0,\mathbf y)$ and $x$ has the
previous form. In this case, $(\Im\Lambda z)^2=
-x^2-(y^2)^2-(y^3)^2\geq -\mathbf y^2-x^2$ and we  see
that~\eqref{4.5} holds again. Finally, let $y$ be lightlike. By
using a real Lorentz transformation, it can be made such that
${\bf y}^2<\epsilon^2$, where $\epsilon$ is arbitrarily small.
Then $(\Im\Lambda z)^2> -\epsilon^2-x^2$ and $\Im\Lambda z\in
\oV^-$ if $\epsilon<l$. Hence, in this case too, $z=x+iy$ is a
point of analyticity at which the function $\Delta_+(z)$
satisfies~\eqref{4.5}. The lemma is proved.

\medskip

{\bf Theorem~3.} {\it Let $W(\xi_1,\dots, \xi_{n-1})$ be the
Wightman function of $\varphi=:\exp g\phi^2:$ defined
by~\eqref{2.4} and \eqref{2.*}. Then, every functional~\eqref{4.4}
extends continuously to the space $A_{2\ell}(V_{(k)})$, where
$\ell=\sqrt{g/6}$ and $V_{(k)}=\{\xi\in\oR^{4(n-1)}\colon \xi_k^2>
0\}$.

\medskip

Proof.} We note that
\begin{equation}
\langle \Psi_0,\, \varphi(x_1)\ldots\varphi(x_{k+1})
\varphi(x_k)\ldots\varphi(x_n)\Psi_0\rangle =\sum_R
\frac{D_R}{R!}\,\check{w}^R,
 \label{4.6}
\end{equation}
where  $\check{w}_{ij}$ is the previous system~\eqref{*} except
for $\check{w}_{k,k+1}=\Delta_+(x_{k+1}-x_k)$. Indeed, let $\tau$
denote the transposition $(1,\dots, k,k+1,\dots, n)\to (1,\dots,
k+1,k, \dots, n)$. Define $R'$  by $r'_{i,j}=r_{\tau i, \tau j}$
for $i\ne k$, $j\ne k+1$ and $r'_{k,k+1}=r_{k,k+1}$. Clearly, we
have the equality
$$
 \prod_{i< j}w_{\tau i\,\tau\! j}^{r'_{ij}}=\prod_{i< j}
 \check w_{ij}^{r_{ij}}.
 $$
The map of multi-indices  $R\to R'$ is bijective. Furthermore,
$R'!=R!$ and $D_{R'}=D_R$ because  $R'_j=R_{\tau j}$.
Thus~\eqref{4.6} does hold. Since the analytic function
$\Delta_+(z)$ is invariant under the total reflection, the
distribution $\check{w}^R(x)$ is a boundary value of the same
holomorphic function  $w^R(z)$ but from the cone
$$
\{y=\Im z\in \oR^n\colon y_i- y_j\in \oV^-,\,\,1\leq i<j\leq
n,\,i\ne k, j\ne k+1;\quad y_k-y_{k+1}\in \oV^+\}.
$$
Let $0<l<\ell$ and let $\eta$ be a vector in  $\oR^{4(n-1)}$ such
that $\eta_j=(-l, 0,0,0)$ for  $j< k-1$, $j>k+1$ and
$\eta_{k-1}=\eta_{k+1}=(-2l,0,0,0)$, $\eta_k=(l, 0,0,0)$. Then all
the differences $y_i- y_j=\eta_i+\dots+\eta_{j-1}$, $i<j$, are in
the suitable cones and hence the value of the distribution
$\check{w}^R(\xi)$ at a test function $f\in
A_{2\ell}(\oR^{4(n-1)})$ can be written as
\begin{equation}
 (\check{w}^R,f)=\int w^R(\xi+i\eta)\,f(\xi+i\eta)\, {\rm
  d}\xi,
  \label{4.7}
 \end{equation}
Using  the invariance of $\Delta_+(z)$ under the total reflection
and the estimate~\eqref{2.2}, we obtain
\begin{equation}
|\check{w}^R(\xi+i\eta)|\leq
 \frac{1}{(2\pi l)^{2|R|}}\,\prod_{i<j}\frac{1}{n_{ij}^{2r_{ij}}}\,,
\label{4.8}
\end{equation}
 where
  $$n_{ij}= |\eta_i^0+\dots+\eta_{j-1}^0|/l
  $$
Next we apply the trick used in the proof of Theorem~1, squaring
the expression~\eqref{4.8} and passing to the symmetric matrix
$(r_{ij})$. Using~\eqref{2.7}, \eqref{2.8}, and the multinomial
theorem, we obtain
\begin{equation}
  \left|\frac{D_R}{R!}\,\check{w}^R(\xi+i\eta)\right|^2 \leq
    \frac{(2g)^{2|R|}}{(2\pi l)^{4|R|}}\,
 \prod_{j=1}^n\left(\sum_{\substack{i=1\\i\ne j}}^n
\frac{1}{n_{ij}^2}\right)^{R_j},
  \label{4.9}
  \end{equation}
where $n_{ij}=|\eta_j^0+\dots+\eta_{i-1}^0|/l$ for $i>j$. It is
easy to see that, for every fixed $j$, the system of nonzero
integers $n_{ij}$, where $1\leq i\leq n$, $i\ne j$, contains at
most two identical integers.  Therefore, we arrive at the
inequality
\begin{equation}
  \left|\frac{D_R}{R!}\,(\check{w}^R,f)\right| \leq C\|f\|_{2l,4n-3}
  \left[\frac{g}{(\pi
l)^2}\sum_{k=1}^{n-1}\frac{1}{k^2}\right]^{|R|}.
  \label{4.10}
  \end{equation}
Taking $l$ sufficiently close to $\ell$, we   infer that the
series $\sum_R(D_R/R!)\,(\check{w}^R,f)$ converges absolutely for
every $f\in A_{2\ell}(\oR^{4(n-1)})$ and defines a continuous
linear functional on this space.

Now let $\eta_k(t)= t(l, 0,0,0)$, where $-1\leq t \leq 1$, and let
the other $\eta_j(t)$, $j\ne k$, be as before. From the above, it
is clear that the value of the functional~\eqref{4.4} at a test
function $f\in A_{2\ell}(\oR^{4(n-1)})$ can be written as
 \begin{equation}
 \sum_R\frac{D_R}{R!}\int_{\oR^{4(n-1)}} \left[w^R(\xi+i\eta(-1))\,f(\xi+i\eta(-1))-
 w^R(\xi+i\eta(1))\,f(\xi+i\eta(1))\right] {\rm
  d}\xi.
  \label{4.11}
 \end{equation}
In extending continuously this functional to $A_{2\ell}(V_{(k)})$,
there is no problem with the contribution to the integral from the
region
\begin{equation}
V^l_{(k)}=\{\xi\in\oR^{4(n-1)}\colon \xi_k^2>-l^2\},
 \notag
 \end{equation}
because the surfaces $V^l_{(k)}+i\eta(\pm1)$ lie in the complex
$2\ell$-neighborhood of the wedge $V_{(k)}$. The surfaces
$\complement V^l_{(k)}+i\eta(\pm1)$, together with
$$
\sigma^l_{(k)}=\{\zeta\in \oC^{4(n-1)}\colon (\Re
\zeta_k)^2=-l^2,\, \Im\zeta=\eta(t), -1\leq t\leq 1\},
$$
compose the piecewise smooth boundary of a $(4n-3)$-dimensional
flat surface in $\oC^{4(n-1)}$ which, by Lemma, is contained in
the analyticity domain of $w^R(\zeta)f(\zeta)$. Considering that
$w^R(\zeta)$ is bounded  and   $f(\zeta)$ decreases rapidly as
$|\Re \zeta|\to \infty$ and applying the Stokes theorem to the
closed form
$\omega^R_f=w^R(\zeta)f(\zeta)d\zeta_1\wedge\dots\wedge
d\zeta_{n-1}$, we obtain
\begin{equation}
\int_{\complement V^l_{(k)}}
\left[w^R(\xi+i\eta(-1))\,f(\xi+i\eta(-1))-
 w^R(\xi+i\eta(1))\,f(\xi+i\eta(1))\right] {\rm d}\xi=
  \int_{\sigma^l_{(k)}}\omega^R_f,
  \notag
 \end{equation}
where the surface $\sigma^l_{(k)}$ is properly oriented. More
explicitly, using the local coordinates $\check\xi=(\xi_1,\dots
\xi_{k-1},\xi_{k+1}\dots\xi_{n-1})$,
$\boldsymbol{\xi}_k=(\xi^1_k,\xi^2_k, \xi^3_k)$, and $t$, we have
$$
\int_{\sigma^l_{(k)}}\omega^R_f=I_++I_-, $$
where
\begin{equation}
|I_\pm|= l\left|\int_{-1}^1{\rm d}t\int_{\oR^{4(n-2)}}{\rm
d}\check\xi\int_{\boldsymbol{\xi}^2>l^2}
w^R(\xi+i\eta(t))\,f(\xi+i\eta(t))
\bigr|_{\xi_k^0=\pm\sqrt{\boldsymbol{\xi}_k^2-l^2}}
 \,{\rm
d}\boldsymbol{\xi}_k\right|.
   \label{4.12}
 \end{equation}
By Lemma,
\begin{equation}
|w_{k,k+1}(\xi,t)|=|\Delta_+(\xi_k+i\eta_k(t))|\leq \frac{1}{(2\pi
l)^2}\quad \text{for} \quad \xi_k^2=-l^2, \, -1\leq t\leq 1.
 \label{4.13}
 \end{equation}
For all other $i<j$, the vector $\eta_i(t)+\dots+\eta_{j-1}(t)$
lies in $\oV^-$ and
\begin{equation}
|w_{ij}(\xi,t)|=|\Delta_+(\xi_i+\dots
+\xi_{j-1}+i\eta_i(t)+\dots+i\eta_{j-1}(t))|\leq \frac{1}{(2\pi
l)^2}\cdot\frac{1}{n_{ij}^2}, \quad -1\leq t\leq 1,
 \label{4.14}
 \end{equation}
because the minimum of $|\eta^0_i(t)+\dots+\eta^0_{j-1}(t)|$
occurs at $t=1$. From~\eqref{4.13} and \eqref{4.14}, it follows
that
\begin{equation}
\left|\int_{\sigma^l_{(k)}}\omega^R_f\right|\leq  C'
\|f\|_{V_{(k)},2l,4n-3}\, \frac{1}{(2\pi
l)^{2|R|}}\,\prod_{i<j}\frac{1}{n_{ij}^{2r_{ij}}}, \label{4.15}
\end{equation}
where $C'=2\int_{\oR^{4(n-2)}}(1+|\xi|)^{-4n+7}{\rm
d}\check\xi\int_{\oR^3} (1+|\boldsymbol{\xi}_k|)^{-4}{\rm
d}\boldsymbol{\xi}_k$. Therefore, an analog of the
inequality~\eqref{4.10} with $\|f\|_{V_{(k)},2l,4n-3}$ instead of
$\|f\|_{2l,4n-3}$ holds for
$(D_R/R!)\int_{\sigma^l_{(k)}}\omega^R_f$ as well as for
$(D_R/R!)\int_{V^l_{(k)}+i\eta(\mp 1)}\omega^R_f$. We conclude
that the series
$$
\sum_R\frac{D_R}{R!}\left(\int_{V^l_{(k)}+i\eta(-1)}
+\int_{V^l_{(k)}+i\eta(1)}+\int_{\sigma^l_{(k)}}\right)\omega^R_f
$$
converges absolutely for any   $f\in A_{2\ell}(V_{(k)})$ and
defines a continuous linear functional which coincides
with~\eqref{4.4} on $A_{2\ell}(\oR^{4(n-1)})$. Theorem~3 is thus
proved.

\section{\large Wightman-type axioms for quasilocal fields}

In this section we discuss general characteristic properties of
the vacuum expectation values in the nonlocal QFT with an
elementary length introduced via the presheaf of spaces
$A_\ell(O)$. For simplicity, we restrict our consideration to the
case of a neutral scalar field $\varphi$. Its  $n$-point vacuum
expectation value  $\langle \Psi_0,\, \varphi(x_1)\ldots
\varphi(x_n)\Psi_0\rangle$ is denoted by ${\mathcal W}_n(x_1,\dots
x_n)$. We hold the usual assumption that these generalized
functions have no worse than polynomial growth in position space
or, what is the same, that their Fourier transforms $\hat{\mathcal
W}_n$ have finite order of singularity. We take the test function
space $A_\infty$ to be the initial  domain of definition of the
field, just as  was done in Refs.~\cite{IF1,IF2}, where, however,
a different characterization of this space was used.  The proper
orthochronous Poincar\'e group ${\mathcal P}^\uparrow_+$ acts on
the space $A_\infty(\oR^{4n})$ by
$$
f_{(a,\Lambda)}(x_1,\dots,x_n)=
f(\Lambda^{-1}(x_1-a),\dots,\Lambda^{-1}(x_n-a)),\qquad
(a,\Lambda)\in {\mathcal P}^\uparrow_+.
$$
We assume that the vacuum expectation values have the following
characteristic properties.

\begin{enumerate}
\item[($a.1$)] ({\it Initial  domain of definition})
$$
{\mathcal W}_n\in A'_\infty(\oR^{4n})\qquad \mbox{for}\quad n\geq
1.
$$

\item[($a.2$)] ({\it Hermiticity})
$$
\overline{({\mathcal W}_n, f)}= ({\mathcal W}_n, f^\dagger)\quad
\mbox{for each}\,\, f\in A_\infty(\oR^{4n}),\quad \mbox{with}\,\,
f^\dagger(x_1,\dots,x_n)=\overline{f(x_n,\dots,x_1)}.
$$

\item[($a.3$)] ({\it Positive definiteness})
\begin{equation}
\sum_{k,m=0}^N ({\mathcal W}_{k+m},f^\dagger_k\otimes f_m)\geq 0,
 \notag
 \end{equation}
where ${\mathcal W}_0=1$, $f_0\in \oC$, and $\{f_1,\dots, f_N\}$
is   an arbitrary  finite set of test functions such that $f_k\in
A_\infty(\oR^{4k})$, $k=1,\dots, N$.

\item[($a.4$)] ({\it Poincar\'e covariance})
$$
({\mathcal W}_n, f)=({\mathcal W}_n, f_{(a,\Lambda)})\qquad
\mbox{for each}\quad f\in A_\infty(\oR^{4n})\quad  \mbox{and}
\quad \mbox{for each} \quad (a,\Lambda)\in {\mathcal
P}^\uparrow_+.
$$
This  is equivalent to the existence of Lorentz-invariant
functionals $W_n\in A'_\infty(\oR^{4(n-1)})$ such that
\begin{equation}
{\mathcal W}_n(x_1,\dots x_n)=W_n(x_1-x_2,\dots,x_n-x_{n-1}),
\quad n\geq1.
 \notag
\end{equation}

\item[($a.5)$] ({\it Spectral condition})
$$
\supp \hat{W}_n\subset
\underbrace{\bar{\oV}^+\times\dots\times\bar{\oV}^+}_{(n-1)}.
$$

\item[($a.6$)] ({\it Cluster decomposition property})\\
If $a$ is a spacelike vector, then for each $f\in
A_\infty(\oR^{4k})$ and for each $g\in A_\infty(\oR^{4m})$,
$$
({\mathcal W}_{k+m},f\otimes g_{(\lambda a,I)})\longrightarrow
({\mathcal W}_k,f)({\mathcal W}_m,g) \qquad \mbox{as}\quad
\lambda\to \infty.
$$

\item[($a.7.1$)] ({\it Quasilocalizability})\\
There exists $\ell<\infty$ such that every functional $W_n$ has a
continuous extension to the space $A_\ell(\oR^{4(n-1)})$.

  \item[($a.7.2$)] ({\it Quasilocality})\\
  For any $n\geq2$ and $1\leq k\leq n-1$, the difference
\begin{multline}
W_n(\xi_1,\dots,\xi_{k-1},\xi_k,\xi_{k+1},\dots, \xi_{n-1})\\ -
W_n(\xi_1,\dots,\xi_{k-1}+\xi_k,-\xi_k,\xi_k+\xi_{k+1},\dots,
\xi_{n-1})
 \label{5.1}
 \end{multline}
(with the above specification for $k=1, n-1$) has a continuous
extension to the space $A_\ell(V_{(k)})$, where
$V_{(k)}=\{\xi\in\oR^{4(n-1)}\colon \xi_k^2> 0\}$.

\end{enumerate}
The listed properties of vacuum expectation values are different
from those of the Wightman functions in local QFT~\cite{SW,J,BLOT}
in two aspects. First, the space $A_\infty$ substitutes for the
Schwartz space $S$ of  infinitely differentiable functions of fast
decrease. It should be emphasized that the role of $A_\infty$ is
completely auxiliary because the true  domain of definition of the
vacuum expectation values is specified by  condition ($a.7.1$).
This condition means, intuitively, that the correlation functions
are localizable with respect to the relative coordinates at scales
larger than $\ell$. The reason is that all observable quantities
are expressed in these variables. The space $A_\infty$ may be
replaced with other test function space which is invariant under
linear transformations of the coordinates and is dense in every
$A_\ell$, $\ell<\infty$, and for which a kernel theorem holds. For
instance, we may use $S^0=\hat{\mathcal D}$ and this gives a
system of axioms which is completely equivalent to the system
listed above. Such a choice is quite convenient because the space
$\mathcal D$ is customary for physicists theoreticians and
occupies a prominent place in functional analysis. On the other
hand, $A_\infty$ is maximal among the spaces  suitable for the
role of an initial  domain of definition of fields in nonlocal
quantum field theory. The second and main difference of our
approach from the traditional formalism~\cite{SW,J,BLOT} of local
QFT is that the local commutativity axiom is replaced with the
weaker quasilocality condition ($a.7.2$) which means, intuitively,
that causality principle is obeyed at spacetime scales large
compared to the fundamental length $\ell$.

Although the space $A_\ell(\oR^{4(n-1)})$ is not Lorentz
invariant,  conditions ($a.7.1$) and ($a.7.2$) can be given an
 invariant form~\cite{S80}. By  condition ($a.4$), the functional $W_n$
is well defined on every space obtained from
$A_\ell(\oR^{4(n-1)})$ by a Lorentz transformation. All the spaces
$\Lambda A_\ell(\oR^{4(n-1)})$, $\Lambda\in L^\uparrow_+$, are
contained in $S^1(\oR^{4(n-1)})$ and their linear span can be
given the inductive limit topology induced by the injections
$\Lambda A_\ell(\oR^{4(n-1)})\to S^1(\oR^{4(n-1)})$. Then $W_n$
extends continuously to the resulting Lorentz invariant space.
Analogously,  condition ($a.7.2$), together with condition
($a.4$), implies that the functional~\eqref{5.1}  has a continuous
extension  to the linear span in $S^1(V_{(k)})$ of the spaces
$\Lambda A_\ell(V_{(k)})$, $\Lambda\in  L^\uparrow_+$, and we
thereby obtain a relativistically invariant implementation of
causality.

In the original Wightman's formulation~\cite{SW}, local
commutativity is expressed as a  property of the distributions
$\mathcal W_n(x_1,\dots,x_n)$ with respect to the permutations of
spacelike separated arguments  $x_k$. For comparison, let us
derive the corresponding consequence of condition  ($a.7.2$). Let
$\pi$  be a permutation of the  indices $1,\dots,n$. It acts on
$\oR^{4n}$  by the rule
$$
\pi(x_1,\dots, x_n)=(x_{\pi^{-1}(1)},\dots, x_{\pi^{-1}(n)})
$$
and acts on  $f(x)$ in an analogous manner  $(\pi
f)(x)=f(\pi^{-1}x)$. We let $\pi\mathcal W_n$ denote the
``permuted'' vacuum expectation value
$\langle\Psi_0,\varphi(x_{\pi 1})\cdots \varphi(x_{\pi
n})\Psi_0\rangle$. In more exact terms,
$$
\left(\pi\mathcal W_n, f\right)=(\mathcal W_n,\pi^{-1}f),\qquad
f\in A_\infty(\oR^{4n}).
$$

\medskip

{\bf Theorem~4.} {\it Condition $(a.7.1)$ implies that for any
$n\geq 1$, the functional ${\mathcal W}_n$ has a continuous
extension to the space $A_{\!\ell_n}(\oR^{4n})$, where
$\ell_n=\ell(n-1)/2$. Furthermore, from  condition $(a.7.2)$ it
follows that for any permutation $\pi$, the difference $\mathcal
W_n-\pi\mathcal W_n$ extends continuously to the space
$A_{\ell_n}(V_\pi)$, where
 \begin{equation}
V_\pi=\bigcup_{k,m} V_{(k,m)},\qquad  V_{(k,m)}=\{x\in
\oR^{4n}\colon (x_k-x_m)^2> 0\},
 \label{5.2}
\end{equation}
and the union ranges over all pairs of indices such that $k<m$ and
$\pi^{-1} k>\pi^{-1} m$.

\medskip

 Proof.} The first statement was already proved in deriving
Corollary  of Theorem~2 and it suffices to verify the second
statement for a transposition $\pi_k\colon
(1,\dots,k,k+1,\dots,n)\to (1,\dots,k+1,k,\dots,n)$. If  $f\in
A_\infty(\oR^{4n})$, then $(\mathcal W_n-\pi_k\mathcal W_n,f)$ is
equal to  the value of  functional~\eqref{5.1} at the test
function $f_t(\xi)=\int_{\oR^4} f(t^{-1}(\xi,X))\,{\rm d}X$, where
$t$ is defined by~\eqref{3.6}. Let us show that $f\to f_t$ is a
continuous map from $A_{\ell_n}(V_{(k,k+1)})$ into
$A_{\ell}(V_{(k)})$. Let $l<\ell$ and let $\zeta\in \tilde
V^l_{(k)}$, i.e., there exists $\bar{\xi}\in V_{(k)}$ such that
$|\zeta-\bar{\xi}|<l$. Suppose that $\bar{x}$ is related to
$\bar{\xi}$ by~\eqref{3.7} and  $z$ is related to $\zeta$  in a
similar way. Clearly, $\bar{x}\in V_{(k,k+1)}$ and, for every
$j=1,\dots,n$, we have
$$
|z_j-\bar{x}_j|\leq\frac{1}{n}\left(\sum_{m=1}^{j-1}m|\zeta_m-\bar{\xi}_m|+
\sum_{m=1}^{n-j}m|\zeta_{n-m}-\bar{\xi}_{n-m}|\right)<
l\frac{n-1}{2}=l_n.
$$
Thus, the point $t^{-1}(\zeta,X)$ belongs to $\tilde
V^{l_n}_{(k,k+1)}$ for any $X\in\oR^4$. This enables one to obtain
an estimate analogous to~\eqref{3.9}, namely,
$$
\|f_t\|_{V_{(k)},l,N}\leq C\|f\|_{V_{(k,k+1)},l_n,N+5},
$$
which completes the proof.

Theorems~2 and 3 show that the vacuum expectation values
calculated from the formula~\eqref{2.6} satisfy  conditions
($a.7.1$) and ($a.7.2$). Deriving these theorems, we used not the
explicit form of the function $:\exp g\phi^2:$ but the
restriction~\eqref{2.8} on the determining coefficients. For this
reason  a   more general statement is true.

\medskip

{\bf  Theorem~5.} {\it Let $\phi$ be a free neutral scalar field
in  Minkowski space and let
\begin{equation}
\varphi(x)= \sum_{r=0}^\infty\frac{d_r}{r!}:\phi^r:(x),
\notag
\end{equation}
where the coefficients $d_r$ are real and subject to the condition
\begin{equation}
d_r^2\leq C (2g)^r r!.
 \notag
\end{equation}
Then, the vacuum expectation values of   $\varphi(x)$ have all
properties $(a.1)-(a.7)$, and the conditions $(a.7.1)$ and
$(a.7.2)$ are satisfied with  $\ell=2\sqrt{g/6}$.

\medskip

Proof.} Conditions ($a.7.1$) and  ($a.7.2$) are fulfilled by
Theorems~2 and 3 and the first of them even with the constant
$\sqrt{g/6}$. Condition ($a.1$) is obviously satisfied by
Corollary of Theorem~2 because $A_\infty$ is contained in every
$A_{\!\ell_n}$. The Hermiticity property ($a.2$) is obvious from
the reality of $d_r$. The positive definiteness ($a.3$) is a
direct consequence of the fact that the Wightman functions of the
Wick polynomials $\sum_{r=0}^N(d_r/r!):\phi^r:(x)$ have this
property, which, in turn, is apparent from the definition of the
Wick monomials $:\phi^r:(x)$ via a limiting procedure. Property
($a.4$) follows immediately from the Lorentz invariance of
$\Delta_+(x)$. The spectral condition ($a.5$) is fulfilled by the
Paley-Wiener-Schwartz theorem (see, e.g., Theorem~2.8 in
Ref.~\cite{SW} or Theorem~B.8 in Ref.~\cite{BLOT}) which shows
that every term of the series defining $\hat{W}_n$ has support in
$\bar{\oV}^+\times\dots\times\bar{\oV}^+$. Hence, the sum of this
series also has this support property because
$\hat{A}_\infty\subset \mathcal D$. It remains to show that the
cluster decomposition property~($a.6$) also holds. We will prove
even a  stronger version of this property.

Let $a$ be a spacelike vector in $\oR^4$  and let $h$ be an
arbitrary element of $A_\infty(\oR^{4n})$, where $n=k+m$. We set
$$
h_{\lambda a}(x_1,\dots,x_k,x_{k+1},\dots,
x_n)=h(x_1,\dots,x_k,x_{k+1}-\lambda a,\dots, x_n-\lambda
a),\qquad \lambda>0.
$$
and claim that the vacuum expectation values of $\varphi(x)$
satisfy the estimate
\begin{equation}
|(\mathcal W_{k+m}-\mathcal W_k\otimes\mathcal W_m,\, h _{\lambda
a})| \leq C_a\|h\|_{n\ell,4(n+1)}\frac{1}{1+\lambda^2},
\label{5.3}
\end{equation}
where the norm of $h$ is defined by~\eqref{3.1}. Setting
$h=f\otimes g$, we see that~\eqref{5.3} implies  ($a.6$) and
moreover the limit converges no worse than $1/\lambda^2$. To
prove~\eqref{5.3}, we note that
\begin{equation}
\mathcal W_{k+m}-\mathcal W_k\otimes\mathcal W_m=
{\sum_R}^\prime\, \frac{D_R}{R!}\,w^R,
 \label{5.4}
\end{equation}
where the prime indicates that the summation ranges over all
matrices $R$ with the property that $r_{ij}\ne0$ for at least one
pair  of indices such that $i\leq k$, $j>k$.  Let $(i_0,j_0)$ be
such a pair and let
\begin{equation}
U_{i_0j_0}=\sum_{\{R\colon r_{i_0j_0}\ne0\}} \frac{D_R}{R!}\,w^R.
 \notag
\end{equation}
We estimate this expression,  setting $\Im z_j=((j-1)\,l,0,0,0)$
for all $j=1,\dots, n$ and assuming that the Lorentz square of
$x_{i_0}-x_{j_0}=\Re(z_{i_0}-z_{j_0})$ is less than
 $-L^2$.   By  Lemma in Sec.~IV, we have
\begin{equation}
|\Delta_+(z_{i_0}-z_{j_0})|\leq
\frac{1}{2\pi^2}\cdot\frac{1}{l^2(j_0-i_0)^2+L^2}
\leq\frac{2n^2l^2}{l^2+L^2}\cdot
\frac{1}{4\pi^2l^2}\cdot\frac{1}{(j_0-i_0)^2}.
 \label{5.5}
\end{equation}
Let $l>\ell$ and $l^2+L^2> 2n^2l^2$. We proceed as in the proof of
Theorem~1, using~\eqref{5.5} and~\eqref{2.9*} for other pairs of
indices, and conclude that for $(x_{i_0}-x_{j_0})^2\leq -L^2$,
\begin{equation}
|U_{i_0j_0}(z)|\leq C_0\frac{2n^2l^2}{l^2+L^2},
 \label{5.6}
\end{equation}
where $C_0$ is a constant dominating the total sum $\sum_R
|D_R\,w^R|/R!$.

With $y=\Im z$ taken as stated above,  we have the representation
\begin{equation}
(U_{i_0j_0},h_{\lambda a} )=I_<+I_>,\quad \mbox{where}\,\,
I_\lessgtr=\int_{(x_{i_0}-x_{j_0})^2\lessgtr-\lambda^2a^2/2}
U_{i_0j_0}(x+iy)h_{\lambda a}(x+iy) {\rm d} x.
 \notag
\end{equation}
Choosing $l$ so that $\ell<l<\ell n/(n-1)$ and using~\eqref{5.6}
with $L^2=\lambda^2a^2/2$, we obtain
\begin{equation}
|I_<|\leq C_1\|h\|_{n\ell,4n+1}\frac{1}{1+\lambda^2}.
 \notag
\end{equation}
From similarity considerations, it is clear that the distance of
the point $-\lambda a$ from the set $\{x\in\oR^4\colon x^2\geq
-\lambda^2 a^2/2\}$ increases linearly with $\lambda$. Because of
this, there is $\epsilon_a>0$ such that $|x_{i_0}-x_{j_0}+\lambda
a|>\epsilon_a\lambda$ for all points in the range of integration
for $I_>$. Therefore,
\begin{multline}
|I_>|\leq C_0\int_{|x_{i_0}-x_{j_0}+\lambda
a|>\epsilon_a\lambda}|h_{\lambda a}(x+iy)|{\rm d}
x=C_0\int_{|x_{i_0}-x_{j_0}|>\epsilon_a\lambda}|h(x+iy)|{\rm d} x\\
\leq C_0\int_{|x_{i_0}-x_{j_0}|>
\epsilon_a\lambda}\frac{\|h\|_{n\ell,N}}{(1+|x|)^N}\, {\rm d}
x\qquad \mbox{for any}\quad N\geq 4n+1.
 \label{5.8}
\end{multline}
Using the elementary inequality $|x_{i_0}-x_{j_0}|\leq 2|x|$ and
setting $N=4(n+1)$, we continue the estimate~\eqref{5.8} as
follows
\begin{equation}
|I_>|\leq
C_0\int_{|x_{i_0}-x_{j_0}|>\epsilon_a\lambda}\frac{8\|h\|_{n\ell,4(n+1)}}
{(2+|x_{i_0}-x_{j_0}|)^3(1+|x|)^{4n+1}}{\rm d}x \leq
C_2\|h\|_{n\ell,4(n+1)}\frac{1}{1+\lambda^3}.\notag
\end{equation}
We see that the term $I_>$ is negligible  compared to $I_<$ and
\begin{equation}
|(U_{i_0j_0},h_{\lambda a} )|\leq
C'_2\|h\|_{n\ell,4(n+1)}\frac{1}{1+\lambda^2}.
 \notag
\end{equation}
Next we choose  another pair of indices $(i_1,j_1)$ such that
$i_1\leq k$, $j_1>k$ and consider
\begin{equation}
U_{i_1j_1}=\sum_{\{R\colon r_{i_1j_1}\ne0, r_{i_oj_0}=0\}}
\frac{D_R}{R!}\,w^R.
 \notag
\end{equation}
Using the same line of reasoning, we conclude that
$(U_{i_1j_1},h_{\lambda a} )$ satisfy an analogous bound. After a
finite number of  steps we exhaust the sum on the right-hand side
of~\eqref{5.4} and arrive at~\eqref{5.3}, which completes the
proof.

Thus the nonlocal field $:\exp g\phi^2:(x)$ completely fits into
the framework proposed in Refs.~\cite{IF1,IF2,FS1}. It should be
stressed that the theory defined by  assumptions ($a.1$) - ($a.7$)
becomes local in the limit  $\ell\to0$. The injective limit
$\injlim_{\ell\to 0}A_\ell(\oR^d)$ coincides with the space
$S^1(\oR^d)$, and the notion of support can be correctly defined
for the continuous linear functionals on this space. This can be
done by adapting the definitions used in the Sato-Martineau theory
of hyperfunctions to the case of functionals of polynomial growth
at infinity. The point is that if $K_1$ and $K_2$ are compact
subsets of $\oR^d$, then the spaces $S^1(K_i)=\injlim_{\ell\to
0}A_\ell(K_i)$, $i=1,2$, obey the following structural relation
\begin{equation}
S^1(K_1\cap K_2)=S^1(K_1)+S^1(K_2),
 \label{5.9}
\end{equation}
which is equivalent to the relation
\begin{equation}
S^1(K_1\cap K_2)'=S^1(K_1)'\cap S^1(K_2)',
 \label{5.10}
\end{equation}
for the dual spaces. As shown in Ref.~\cite{S1}, these relations
can be extended to compact sets in the radial compactification
$\hat{\oR}^d$ of $\oR^d$. Because of this, for every  $v\in
S^1(\oR^d)'$, there exists a unique minimal compact set $K\subset
\hat{\oR}^d$ such that $v\in S^1(K)'$. If  Wightman functions
$\mathcal W_n$ satisfy  conditions ($a.7.1$) and ($a.7.2$) for
each $\ell>0$, then from Theorem~4 it follows that they are
defined on the spaces $S^1(\oR^{4n})$. Moreover, then all axioms
($a.1$) - ($a.6$) are fulfilled with test functions of the class
$S^1$ because $A_\infty(\oR^d)$ is dense in $S^1(\oR^d)$.
Theorem~4 also implies that   $\mathcal W_n-\pi\mathcal W_n$
admits a continuous extension to the space $S^1(V_\pi)$ in this
limit, i.e., the quasilocality condition ($a.7.2$) turns into
local commutativity understood in the sense of hyperfunctions.

\medskip

{\large Remark~5.} In the axiomatic scheme proposed in~\cite{BN},
the role of condition~($a.7.1$) is played by an assumption which
is equivalent to saying that every functional $W_n$ has a
continuous extension to each of the spaces~\eqref{1.2}, where the
tensor product is endowed with the projective topology. This
assumption allows an exponential growth of order~1 and of
arbitrary  type when more than one of the momentum-space variables
$q_j$ tend to infinity together. Because such is the case for any
$\ell$, it is unlikely that the scheme~\cite{BN} becomes local in
the limit $\ell\to0$. In place of condition ($a.7.2$), a weaker
condition was used in~\cite{BN}, namely, that the
functionals~\eqref{4.1} extend continuously to the spaces denoted
there by $\mathcal T(W_k^{\ell'})$. In our notation, this means
that there exist continuous extensions to
$A_\infty(\oR^{4(k-1)})\otimes A_\ell(V_{(k,k+1)})\otimes
A_\infty(\oR^{4(n-k-1)})$, but this condition does not turn into
local commutativity as $\ell\to0$. In particular, there is
seemingly no analogue of Theorem~4.

\section{\large Conclusion}

The analysis performed in this paper supports the view~\cite{FS1}
that the presheaf of spaces $A_\ell(O)$ associated with sets
$O\subset \oR^d$ provides a suitable framework for constructing
quantum field theory with a fundamental length. Our approach
agrees with the idea that localizability at large scales should be
understood as a property of the correlation functions with respect
to the relative coordinates, because observable quantities are
expressed in these variables. It also meets the requirement that
the theory must become local as the fundamental length tends to
zero. The space $A_\infty(\oR^d)$ corresponding to the
ultra-hyperfunctions can be used to construct an invariant domain
of  fields in the Hilbert space of states, but it plays  an
auxiliary role.  The situation is somewhat similar to that with
the Wick exponential of the massless scalar  field or of the ghost
field in  a two dimensional spacetime, because the operator
realization of these models is possible only under further
restrictions~\cite{Pi, MS} on test functions besides those
required by the Wightman functions. The reason was there in the
occurrence of infrared singularities and in the lack of
positivity. In the case of nonlocal field $:\exp g\phi^2:$, the
reason is more prosaic and has to do with the fact that
$A_\ell(\oR^d)$ is not invariant under linear transformations of
$\oR^d$. The space $A_\infty(\oR^d)$ is its  maximal invariant
subspace, but one may also use smaller invariant dense subspaces
and $S^0(\oR^d)$ in particular.

If an investigation of physically relevant models related, e.g.,
to string theory will give a good reason,  then conditions
($a.7.1$) and ($a.7.2$) can  be  weakened by assuming that the
nonlocality parameter increases with $n$. Condition ($a.7.2$)
might simply be replaced with the requirement that, for any
permutation $\pi$, the difference $\mathcal W_n-\pi\mathcal W_n$
has a continuous extension to the space $A_{2\ell_n}(V_\pi)$,
where $\ell_n$ grows linearly with increasing $n$. However, in our
opinion, modifications in these conditions may not involve test
functions of the class $A_\infty$ even in respect to a part of
variables because this leads to the lack of the local limit.

Any axiomatic scheme is interesting not by itself but by its
physical consequences. It should be noted in this connection that
the distinction between our formulation and that of~\cite{BN}
becomes quite essential in deriving the analyticity properties of
scattering amplitudes and in finding  bounds on  their high-energy
behavior. Our approach admits an extension to the
Lehmann-Symansik-Zimmermann formalism which play an important role
in the scattering theory. In particular, let $R(x;x_1,\dots,x_n)$
be the retarded product which is formally defined in local QFT by
\begin{multline}
R(x;x_1,\dots,x_n)\\
= (-i)^n\sum_\pi \theta(x^0-x^0_{\pi1})
\theta(x^0_{\pi1}-x^0_{\pi2})\dots\theta(x^0_{\pi(n-1)}-x^0_{\pi
n}) \,[\dots[\varphi(x),\varphi(x_{\pi1})],\dots,\varphi(x_{\pi
n})]. \notag
\end{multline}
Then a natural generalization of its support properties to the
case of a nonlocal field is the requirement of the existence of a
continuous extension to the space $A_{\ell_n}({\mathcal V}^+)$,
where ${\mathcal V}^+=\{(x;x_1,\dots,x_n)\colon (x-x_j)\in
\bar{\oV}^+, j=1,\dots,n\}$.

We conclude by noting that the condition $a.7.2$ is evidently
stronger than the condition that the matrix elements of the
commutator  $[\varphi(x_1),\varphi(x_2)]$ have continuous
extensions to the space $S^0(\bar{V}_{(1,2)})$. The latter is
referred to as asymptotic commutativity~\cite{FS2} and even this
weaker condition ensures the existence of the $CPT$-symmetry (with
respect to the product of charge conjugation $C$, space inversion
$P$, and time inversion $T$) and the normal spin-statistics
connection for nonlocal fields, as has been shown in~\cite{S2}
with the use of the notion of analytic wave front set of a
distribution. Detailed proofs of these theorems for the general
case of a finite family of fields $\{\phi_\iota\}$,
$\iota=1,\dots,I$, transforming according to irreducible
representations of the proper Lorentz group $L_+^\uparrow$ or its
covering group $SL(2,\oC)$ are given in~\cite{S3}.

\medskip
\section* {\large Acknowledgments}
This paper was supported in part by the the Russian Foundation for
Basic Research (Grant No.~09-01-00835) and the Program for
Supporting Leading Scientific Schools (Grant No.~LSS-1615.2008).

\section*{\large Appendix. A density theorem}

As well known, the Fourier transformation maps isomorphically the
space  $\mathcal D(\oR^d)$ of compactly supported smooth functions
on  $\oR^d$ onto the space $S^0(\oR^d)$ consisting of entire
functions on $\oC^d$ with the property that
$$
|z^\kappa g(z)|\leq C_\kappa e^{a|\Im z|},\qquad \kappa\in
\oZ_+^d,
$$
where  $C_\kappa$ and $a$ are constants depending on $g$. The
notation  $S^0(\oR^d)$ indicates that this space is smallest among
the Gelfand-Shilov spaces $S^\alpha(\oR^d)$, $\alpha\geq 0$.

{\bf Proposition.} {\it The space $S^0(\oR^d)$  is dense in
$A_\ell(\oR^d)$ for any $\ell\in(0,+\infty]$.

Proof.}  Since $S^0(\oR^d)$ is an algebra under multiplication, it
contains  a function $g$ such that
$$
g(x)\geq 0\quad  \text{for all}\quad  x\in \oR^d\quad
\text{and}\quad \int g(x){\rm d}\, x=1.
$$
Let $g_\nu(x)=\nu^dg(\nu x)$, $\nu=1,2,\dots$  and let $f\in
A_\ell(\oR^d)$, $f_\nu(z)=\int g_\nu(z-x)f(x){\rm d} x$. The
functions $g_\nu$ and $f$ are analytic in the domain $|\Im
z|<\ell$ and  decrease rapidly as $\Re z\to \infty$. Therefore for
any $z_0$ whose imaginary part  $y_0$ satisfies $|y_0|<\ell$, we
have
$$
f_\nu(z_0)=\int g_\nu(x_0-x)f(x+iy_0)\,{\rm d} x.
$$
It follows that $f_\nu(z_0)\to f(z_0)$ as $\nu\to \infty$. Setting
$l<\ell$ and using the definition~\eqref{3.1} and the inequality
$(1+|x_0|)\leq (1+|x_0-x|)(1+|x|)$,   we obtain
$$
\|f_\nu\|_{l,N}\leq \|f\|_{l,N}\int (1+|x_0-x|)^N
g_\nu(x_0-x)\,{\rm d} x\leq C_N \|f\|_{l,N},
$$
where $C_N=\int(1+|x|)^Ng(x){\rm d} x$. Because every bounded set
in a Montel space is relatively compact, we conclude that
$f_\nu\to f$ in the topology of $A_\ell(\oR^d)$.

\end{document}